\documentclass[prb,twocolumn,aps,showpacs]{revtex4}
\usepackage{graphicx}% complex graphics
\usepackage{bm}% bold math
\usepackage{amssymb} %math symbols
\begin{document}

\title{2D skew scattering in the vicinity and away from resonant
scattering condition}

\author{V. V. Mkhitaryan and M. E. Raikh}

\affiliation{ Department of Physics, University of Utah, Salt Lake
City, UT 84112}

\begin{abstract}
We studied the energy dependence of the 2D skew scattering from
strong potential, for which the Born approximation is not
applicable. Since  the skew scattering cross section is zero both
at low and at high energies, it exhibits a maximum as a function
of energy of incident electron. We found analytically the shape of
the maximum for an exactly solvable model of circular-barrier
potential. Within a rescaling factor, this shape is universal for
strong potentials. If the repulsive potential has an attractive
core, the discrete levels of the core become quasilocal due to
degeneracy with continuum. For energy of incident electron close
to the quasilocal state with zero angular momentum, the
enhancement of the net cross section is accompanied by resonant
enhancement of the skew scattering. By contrast, near the
resonance with quasilocal states having momenta $\pm 1$,  the skew
scattering cross section is an {\em odd} function of energy
deviation from the resonance, and passes through zero, i.e.,
it exhibits a {\em sign reversal}. In the
latter case, in the presence of the Fermi sea, the Kondo resonance
manifests itself in strong temperature dependence of the skew
scattering.
\end{abstract}
\maketitle

\section{Introduction}
The phenomenon of  skew scattering has its origin
in a spin-orbit part
$\propto \hat{\bm{\sigma}}\cdot \left[{\bf k}\times \nabla V(\rho)\right]$
of the scattering potential, $V(\rho)$. Due to this
part, the scattering amplitude between
the states with momenta ${\bf k}$ and ${\bf k}^{\prime}$ acquires a
 contribution,
$f^a\propto \hat{\bm{\sigma}}\cdot\left[{\bf k}\times {\bf k}^{\prime}\right]$,
which is asymmetric
with respect to the scattering angle,
$\theta$, between ${\bf k}$ and ${\bf k}^{\prime}$.
%$\sin \theta = \left({\bf k}\times {\bf k}^{\prime}\right)_z/|{\bf k}|^2$.
Therefore, in two dimensions,  differential cross section acquires
an asymmetric contribution
\begin{equation}
\frac{d\sigma^a}{d\theta}= 2\Re e \left[(f^s)^{\ast}f^a\right]\propto \sigma \sin\theta
\end{equation}
where $f^s$ is the symmetric part of the scattering amplitude, i.e.,  the
scattering amplitude in absence of  spin-orbit potential, and $\sigma=\pm 1$
is the spin projection on the normal to the 2D plane.

Skew scattering,
introduced almost 80 years ago \cite{Mott, Smit},
had recently attracted a lot of interest, since it is a key
ingredient of the anomalous Hall effect \cite{Luttinger, Muttalib1,
Muttalib2, exp1,aronov,Sinitsyn} as well as of the spin-Hall effect
\cite{Dyakonov1, Dyakonov2, Hirsh, Engel, Chudnovsky, Vignale,
Vignale1, Mishchenko, review}. Detailed theoretical calculations of
$\sigma^a$ were carried out for atomic systems \cite{Motz}. Complexity
of these calculations stems from the fact that $\sigma^a\equiv 0$ in the
lowest Born approximation for $f^s$. The results of the second Born
approximation for $f^s$,
yielding a finite $\sigma^a$, are presented in Ref. \onlinecite{Motz}
for the model of the Thomas-Fermi screening of the charge of a nucleus.

Calculations of Ref. \onlinecite{Motz} were recently utilized in
Ref. \onlinecite{Engel} to estimate the
magnitude of skew scattering by a donor impurity in 2D electron gas.
Concerning the energy dependence of the skew-scattering, $\sigma^{a}(E)$,
in calculation of the anomalous Hall and  spin-Hall
effects, it should be taken at $E=E_F$, where $E_F$ is the
Fermi energy. Thus, the skew scattering is sensitive to
the electron density via $E_F$. Then the
question about the explicit form of $\sigma^{a}(E)$
in two dimensions arises.
Nontriviality of this dependence stems from the fact
that at low energy  $f^s(E)$,
is dominated by the angular momentum, $l=0$. As a result,
the skew-scattering cross section, $\sigma^a(E)$,
turns to zero \cite{Mott} at $E\rightarrow 0$. Since $\sigma^a(E)$ turns
to zero at large $E$ as well, we conclude that it should
{\em pass through a maximum} at a certain finite energy.

Study of the energy dependence of the skew scattering is the focus
of the present paper. We demonstrate that this dependence exhibits
especially rich behavior for a ``strong'' scattering potential
with characteristic magnitude, $V_0$, and characteristic radius,
$b$, satisfying the condition $V_0\gg \hbar^2/mb^2$, where $m$ is
the electron mass. For such potentials, the Born approximation
does not apply at low energies. Then the calculation of the skew
scattering requires the knowledge of scattering phases {\em in the
absence of spin-orbit coupling}, while the spin-orbit coupling
should still be treated {\em perturbatively}.

%%%%%%%%%%%%%%%%%%%
\begin{figure}[t]
\centerline{\includegraphics[width=75mm,angle=0,clip]{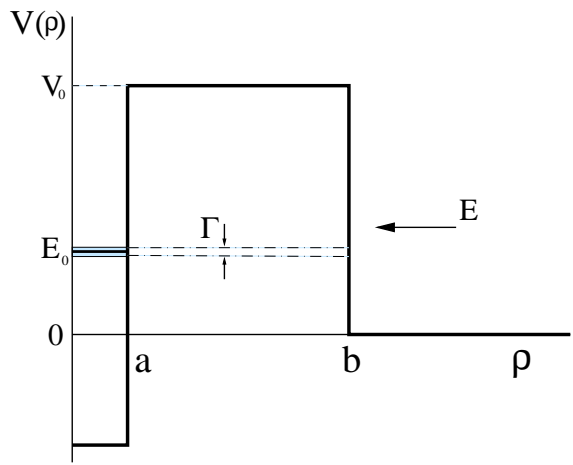}}
\caption{Schematic plot of circular-barrier potential with height,
$V_0$, and radius, $b$. Attractive core at $\rho<a$ contains a
localized state with energy, $E_0$, which has a width, $\Gamma$,
due to degeneracy with continuum.}
\end{figure}
%%%%%%%%%%%%%%%%%%

Even more interesting energy dependence of the skew scattering
emerges in the case when the scattering potential possesses a {\em
core}, as shown in Fig. 1. Then the quasilocal state, $E_0$, due
to this core, has a small width, $\Gamma$. It is known, that
spin-independent cross section exhibits a resonant behavior in the
vicinity  of $E_0$. We will demonstrate that  skew scattering is
also resonantly enhanced near $E=E_0$. A particular interesting
situation realizes when the quasilocal state at $E=E_1$
corresponds to nonzero momenta, $l=\pm 1$. The peculiarity of this
situation is that weak spin-orbit-induced splitting of  the states
with $l=1$ and $l=-1$ can be comparable to $\Gamma$. Then the
dependence, $\sigma^a(E)$, is nonperurbative with respect to
spin-orbit coupling. Moreover, we demonstrate that the skew
scattering  can {\em change sign} within a narrow energy interval
around $E=E_1$. Finally we show, in the presence of the Fermi sea,
the Kondo physics, that modifies the elastic scattering with $l=0$
also modifies the energy dependence of the skew scattering near
the Fermi level within the narrow interval $(E-E_F)$ of the order
of the Kondo temperature.

To the best of our knowledge, the ``exact'' (without utilizing the
Born approximation) energy dependence of the 2D skew scattering
has been  previously studied by Hankiewicz and Vignale
\cite{Vignale} and by  Huang {\em et al.} \cite{Huang} In both
papers the study of  $\sigma^a(E)$ was carried out without
expansion with respect to a weak spin-orbit coupling; the results
for the skew scattering are presented in the form of numerical
plots for certain parameters of the scattering potential and the
magnitude of the spin-orbit coupling. This complicates the
inference of general features of $\sigma^a(E)$. We, on the other
hand, made use of the smallness of the spin-orbit coupling {\em at
the first step}. This allowed us to obtain closed {\em
analytical} results for the skew scattering as a function of
energy. For strong scattering potential, $V(\rho)$, we show that
the energy dependence of $\sigma^{a}$ is universal within two
parametrically wide domains of low energies before the Born
approximation becomes valid. Characteristics of $V(\rho)$
determine the energy scales, but not the form of $\sigma^{a}(E)$
in these domains. Concerning the resonances, around energies
$E=E_0$ and $E=E_1$ of the quasilocal states, $\sigma^{a}(E)$
exhibits a universal behavior as well, in the same way as the
spin-independent scattering. Since quasilocal states manifest
themselves only in the narrow energy intervals around $E=E_0$ and
$E=E_1$, we consider resonant and non-resonant scattering
separately.

The paper is organized as follows. In Sec. II we derive a general
expression for skew-scattering part of the 2D transport scattering
cross section. This expression is linear in spin-orbit coupling
and contains all the  scattering phases in the absence of
spin-orbit coupling. In Sec.~III we extract the energy dependence
of the skew scattering in different domains for a model
circular-barrier potential. Resonant skew scattering is studied in
detail in Sec. IV. In Sec. V. we consider the skew scattering in
the Kondo regime. Concluding remarks are presented in Sec. VI.

\section{General formalism}

As a result of the spin-orbit term
\begin{eqnarray}\label{SO}
{\hat H}_{so}=
\lambda\sigma\frac{1}{\rho}\frac{dV(\rho)}{d\rho}\hat{L}_z
\end{eqnarray}
in the Hamiltonian, where $\lambda$ is the spin-orbit constant and
$\hat{L}_z=-i\hbar\partial/\partial\theta$ is the $z$-component of
the orbital angular momentum, we have $\delta_{l,\sigma}\neq
\delta_{l,-\sigma}$, where $\delta_{l,\sigma}$ are the scattering
phases in the channel with orbital momentum $l$.

Characteristics of the skew scattering, relevant
for transport \cite{Vignale} is
\begin{eqnarray}\label{intrate}
I^a(E)=\frac\sigma\pi\!\int_0^{2\pi}\!\!d\theta
\left(\frac{d\sigma^{a}} {d\theta}\right)\sin\theta,
\end{eqnarray}
where $d\sigma^{a}/d\theta$ is the asymmetric part of the
differential scattering cross section
\begin{eqnarray}
\frac{d\sigma^c(E)}{d\theta}=\frac{d\sigma^{s}(E)} {d\theta}+
\frac{d\sigma^{a}(E)} {d\theta}.
\end{eqnarray}
While $\sigma$ is explicitly present in Eq.~(\ref{intrate}),
$I^a(E)$ is, in fact, $\sigma$-independent. Expression for
$d\sigma^{a}/d\theta$ in terms of scattering phases reads
\begin{eqnarray}\label{rateviaphase}
\frac{d\sigma^{a}}{d\theta} =\frac{i}{2\pi k} \!\sum_{l,l^\prime} (
e^{2i\delta_{l\sigma}}-1)(e^{-2i\delta_{l^\prime\sigma}}-1)
\sin[(l-l^\prime)\theta].
\end{eqnarray}
Eq. (\ref{rateviaphase}) yields for $I^a(E)$ the following general
relation:
\begin{eqnarray}\label{SR}
&&\hspace{-0.3cm}I^{a}(E)=\frac{2\sigma}{\pi
k}\sum_{l=l^\prime\pm1}
\frac{i\,\mbox{sign}(l-l^\prime)}{(\cot\delta_{l,\sigma}-i)(\cot\delta_{l^\prime,\sigma}+i)}\\
&&\hspace{-0.3cm}=\!\frac{\sigma}{\pi k}\!
%\nonumber\\ &&\times
\sum_{l}\!\Bigl\{\sin[2(\delta_{\scriptscriptstyle
l,\sigma}-\delta_{\scriptscriptstyle l+1,\sigma})]
-\sin[2\delta_{\scriptscriptstyle l,\sigma}] + \sin[2\delta_{
\scriptscriptstyle l+1,\sigma}] \Bigr\}.\nonumber
\end{eqnarray}
Obviously, the two last terms of the second line of Eq.~(\ref{SR})
have vanishing contribution, and we will omit them later on.

It is convenient to separate the spin-dependent and
spin-independent parts of scattering phases
\begin{eqnarray}\label{separate}
\delta_{l,\sigma}(E)=\delta^{0}_{l}(E)+
\sigma\delta^{1}_{l}(E).
\end{eqnarray}
Spin-orbit correction, $\delta^{1}_{l}(E)$, can be expressed in
terms of radial eigenfunctions of the continuous spectrum,
$\chi_l(\rho,E)$, in the absence ${\hat H}_{so}$ as follows
\begin{eqnarray}\label{perturbative}
\delta^{1}_{l}(E)=\frac{2m\lambda
l}{\hbar^2}\int_0^{\infty}d\rho\frac1\rho \frac{dV(\rho)}{d\rho}
\chi_l^2(\rho,E).
\end{eqnarray}
Large-$\rho$ behavior of properly normalized $\chi_l(\rho,E)$ is
\begin{eqnarray}\label{normalized}
\chi_l(\rho,E)=\sqrt{\frac{\pi\rho}2}
\Bigl[\cos\delta_l^0(k)J_{|l|}(k\rho)
-\sin\delta_l^0(k)N_{|l|}(k\rho)\Bigr].\nonumber\\
\end{eqnarray}
Substitution of Eq.~(\ref{separate}) into Eq.~(\ref{SR}) and expansion
over $\delta_l^{1}$ yields
\begin{eqnarray}\label{PR}
I^{a}(E)=\frac{2}{\pi k}
\sum_{l}\bigl(\delta^1_{l}-\delta^1_{l+1}\bigr)
\cos2\bigl(\delta^0_{l}-\delta^0_{l+1}\bigr).
\end{eqnarray}
Via simple transformation of the sum in Eq.~(\ref{PR}) and
insertion of the expression for $\delta_l^{1}$,
Eq.~(\ref{perturbative}), we find
\begin{eqnarray}\label{PR1}
I^{a}(E)&=&\frac{4m\lambda }{\pi
k\hbar^2}\sum_{l=-\infty}^{\infty}
\int_0^{\infty}\frac{d\rho}\rho\frac{dV(\rho)}{d\rho}\Bigl[l\chi_{l}^2(\rho,E)
\nonumber\\
&-&(l+1)\chi_{l+1}^2(\rho,E)\Bigr]\cos[2(\delta^0_{l}-\delta^0_{l+1})].
\end{eqnarray}
In Eq.~(\ref{PR1}), it is also convenient to make use of the
identity
\begin{eqnarray}\label{viaDelta}
\cos2(\delta_l^0-\delta_{l-1}^0)=1-2\frac{(\tan\delta_{l}^0-
\tan\delta_{l-1}^0)^2}{(1+\tan^2\delta_{l}^0)
(1+\tan^2\delta_{l-1}^0)}.\nonumber\\
\end{eqnarray}
The advantage of using this identity is that the first term
%in Eq.~(\ref{viaDelta})
does not contribute to the sum in Eq.~(\ref{PR1}), and the
expression for $I^a(E)$ acquires the form
\begin{eqnarray}\label{PR2}
I^{a}(E)&\!\!\!\!=\!\!\!\!&\frac{16m\lambda }{\pi
k\hbar^2}\!\sum_{l=1}^{\infty}
\int_0^{\infty}\frac{d\rho}\rho\frac{dV(\rho)}{d\rho}\Bigl[l\chi_{l}^2(\rho,E)
\\
&&\hspace{-0.4cm}-(l-1)\chi_{l-1}^2(\rho,E)\Bigr]\frac{(\tan\delta_{l}^0-
\tan\delta_{l-1}^0)^2}{(1+\tan^2\delta_{l}^0)
(1+\tan^2\delta_{l-1}^0)}.\nonumber
\end{eqnarray}
The form Eq. (\ref{PR2}) clearly illustrates that,
for a weak scattering potential, skew scattering
is $\propto V^3$. Indeed, when the scattering phases
are small, the last fraction in Eq. (\ref{PR2})
can be replaced by $(\delta_l-\delta_{l-1})^2\propto V^2$,
while for $\chi_l(\rho)$ one can use the free radial
wave functions $\chi_l^0(\rho,E)=\left(\pi\rho/2\right)^{1/2}J_{l}(k\rho)$,
Another power of $V$ comes from $\frac{dV(\rho)}{d\rho}$.
Eq. (\ref{PR2}) is also convenient for the analysis of
the energy dependence of the skew scattering from a strong
potential. We perform this study in the next Section.

\section{Circular-barrier potential}

It is a textbook knowledge \cite{Landau}  that
spin-independent scattering cross section from a
strong potential, $V\gg \hbar^2/mb^2$,
has different forms in three energy intervals:

\noindent (i) low-energy scattering, $E\ll\frac{\hbar^2}{mb^2}$,
in which only zero orbital momentum, $l=0$, contributes to the
cross section;

\noindent (ii) intermediate energies (semiclassical regime),
$\frac{\hbar^2}{mb^2}\ll E \ll  V\left[V/\left(\frac{\hbar^2}{mb^2}\right)\right]$,
where the scattering cross section is determined by high $l\gg 1$,  and finally,

\noindent (iii) high-energy scattering,
$E\gg V\left[V/\left(\frac{\hbar^2}{mb^2}\right)\right]$,
which corresponds to the Born approximation.

As we will see below, the skew scattering exhibits different
universal behaviors in the above three domains. We will see that,
within the interval (ii), in contrast to the spin-independent
scattering, the skew scattering has an additional scale at $E\sim
V_0$, where it passes through the maximum value. Note that, for
strong potential, the energy $E\sim V_0$ is intermediate between
$\hbar^2/mb^2$ and $mV_0^2b^2/\hbar^2$, which are the boundaries
of the interval (ii).

We will perform calculations for a model of circular-barrier
potential $V(\rho)=V_0\theta(b-\rho)$. In principle,  this model
allows to incorporate spin-orbit interaction nonperturbatively
\cite{Vignale}. However, we will use the fact that this
interaction is weak and evaluate Eq. (\ref{PR2}) expanded with
respect to the coupling parameter, $\lambda$.

Within the model $V(\rho)=V_0\theta(b-\rho)$, the phases
$\delta_l^0$ are found from matching at $\rho=b$ the expressions
for $\chi^{\prime}(\rho)/\chi(\rho)$ inside and outside the well.
This yields
\begin{eqnarray}\label{0phase}\tan\delta^0_{l}(k)=
\frac{J^\prime_{|l|}(kb )I_{|l|}(\nu b )- \frac\nu k I^\prime_{|l|}
(\nu b ) J_{|l|}(kb )}{N^\prime_{|l|}(kb )I_{|l|}(\nu b )-
\frac\nu k I^\prime_{|l|}(\nu b ) N_{|l|}(kb )},
\end{eqnarray}
where
\begin{eqnarray}\label{0pa}
\nu=\sqrt{v_0-k^2},\quad v_0= 2m V_0/\hbar^2.
\end{eqnarray}
It is convenient to rewrite Eq. (\ref{0phase}) in a different form
using the recurrence relations for $J_l(z)$ and $N_l(z)$
\begin{eqnarray}\label{0phase1}\tan\delta^0_{l}(k)=
\frac{\left(\frac l{kb}-\frac\nu k\frac{I^\prime_{|l|}(\nu b )}
{I_{|l|}(\nu b )}\right)J_{|l|}(kb )- J_{|l|+1}(kb )}
{\left(\frac l{kb}-\frac\nu k\frac{I^\prime_{|l|}(\nu b )}
{I_{|l|}(\nu b )}\right)N_{|l|}(kb )- N_{|l|+1}(kb )}.
\end{eqnarray}
Also, within a model of a circular well, we have $\frac{dV}{d\rho}
\propto \delta (\rho-b)$, so that the general expression
Eq.~(\ref{PR1}) assumes the form
\begin{eqnarray}\label{final}
&&I^{a}(E)=\frac{4\lambda v_0}k\\
&&\times\sum_{l=1}^{\infty} \frac{(\tan\delta_{l}^0-
\tan\delta_{l-1}^0)^2}{(1+\tan^2\delta_{l}^0)
(1+\tan^2\delta_{l-1}^0)}\left[\frac{l}{Z_l}-\frac{l-1}{Z_{l-1}}\right].\nonumber
\end{eqnarray}
Here we have introduced a notation
\begin{eqnarray}\label{zl}
 Z_{|l|}= \frac{1+\tan^2\delta_l^0(k)}{\bigl[J_{|l|}(kb)
-\tan\delta_l^0(k)N_{|l|}(kb)\bigr]^2},
\end{eqnarray}
so that the spin-orbit corrections to the scattering phases are
given by
\begin{eqnarray}\label{viaz}
\delta^1_l=\frac{\pi\lambda l v_0}{2Z_{|l|}}.
\end{eqnarray}
From Eq.~(\ref{0phase1}), we can express $Z_{|l|}$ via the Bessel
functions
\begin{widetext}
\begin{eqnarray}\label{bigzl}
Z_{|l|}=\left(\frac{\pi kb}2\right)^2\left\{\left[\left(\frac l{kb}-\frac\nu k\frac{I^\prime_{|l|}(\nu b )}
{I_{|l|}(\nu b )}\right)J_{|l|}(kb)-J_{|l|+1}(kb)\right]^2
+\left[\left(\frac l{kb}-\frac\nu k\frac{I^\prime_{|l|}
(\nu b )}
{I_{|l|}(\nu b )}\right)N_{|l|}(kb)-N_{|l|+1}(kb)\right]^2\right\}.
\end{eqnarray}
\end{widetext}

We start the analysis of Eq. (\ref{final}) for the strong
potential $V_0\gg \frac{\hbar^2}{mb^2}$ from the low-energy domain
(i),
 $kb \ll 1$. In this domain, spin-independent scattering is dominated by a single
phase
\begin{eqnarray}\label{delta0}
\delta_0^{0}\approx \arctan\left[\frac{\pi}{2\ln(kb)}\right],
\end{eqnarray}
which follows from $z\ll 1$ behavior of $N_0(z)$. The phases with
higher momenta are much smaller, namely $\delta_l^{0}(E)\sim
(kb)^{2l}$. On the other hand, parameters $Z_l$ grow rapidly with
$l$,
\begin{eqnarray}
\label{ZL} Z_l=v_0b^2\,\frac{\Gamma^2(l)}{\pi^2}
\left(\frac2{kb} \right)^{2l},
\end{eqnarray}
as follows from Eq.~(\ref{bigzl}) (here $\Gamma(l)$ is the Gamma-function).  For this reason
we can replace the sum in Eq.~(\ref{final}) by a single
term, $\bigl(\delta_0^{0}\bigr)^2/Z_1$, so that
\begin{eqnarray}\label{smallkb}
I^{a}(E)\Big |_{\scriptscriptstyle kb\ll 1}\!\!\!=\frac{4\lambda
v_0}k \frac{\bigl(\delta_0^{0}\bigr)^2}{Z_1}
=\left(\!\frac{\pi^2\lambda}
{b}\!\right)\!\frac{kb}{\ln^2(kb)}.\qquad
\end{eqnarray}
First we emphasize that, unlike $\sigma^s$,
the low-energy skew scattering is governed by {\em two}
phases, $\delta_0^{0}$ and $\delta_1^{0}$; the latter phase
enters into $Z_1$, as can be seen from Eq.~(\ref{zl}).
This is a natural consequence of the form Eq. (\ref{SO})
of the spin-orbit Hamiltonian. We also note that,
compared to $\sigma^s$, which behaves as
$\sigma^s(E)\propto 1/\ln^2(kb)$, Eq.~(\ref{smallkb}) contains
an additional factor $(kb)\propto E^{1/2}$, growing rapidly
with $E$. Although derived for particular model of
a circular-well potential, this low-$E$ result is universal
within a factor for a general ``strong'' potential, $v_0b^2\gg 1$.

Next we consider the energy domain $1\ll kb \ll v_0^{1/2}b$, which
belongs to (ii). In this domain, the following simplifications are
possible. Firstly, since $\nu b \approx v_0^{1/2}b\gg kb$, we can
replace $ I^{\prime}_l(\nu b)/I_l(\nu b)$ in Eq.~(\ref{bigzl}) by
$1$. Such a replacement is valid only for $l \ll\nu b$. However,
the relevant range of momenta in this domain is narrower, namely
$l<kb \ll \nu b$. This is because, for $l < kb$, the Bessel
functions, $J_l(kb)$ and $N_l(kb)$ oscillate, while for $l > kb$
they behave as $J_l(kb)\sim(kb/l)^l$ and $N_l(kb)\sim(kb/l)^{-l}$.
This behavior suggests that, $\tan\delta_l^0$ falls off very
rapidly once $l$ exceeds $kb$, and the sum over $l$ in
Eq.~(\ref{final}) should be terminated \cite{plusminus} at $l=kb$.
Note now, that in the interval $1<l<kb$, the ratio $l/(kb)$ in
Eq.~(\ref{0phase1}) can be neglected compared to $\nu/k$. This
yields the following simplified expression for the phases,
$\delta_l^0(E)$,
\begin{eqnarray}\label{asymtan}
\tan \delta_l^0=-\tan\left\{kb+\arctan\left(\frac\nu{k}
\right)-\frac{\pi l}2 -\frac\pi 4 \right\}.\qquad
\end{eqnarray}
On the other hand, neglecting $l/(kb)$ in Eq.~(\ref{bigzl})
leads us to the conclusion that  in the interval $1<l<kb$
all $Z_l$ are equal to each other and are equal to
\begin{eqnarray}\label{Asymz}
Z_l=\frac{\pi kb}{2}\left(1+\frac{\nu^2}{k^2}\right)=\frac{\pi v_0b}{2k}.
\end{eqnarray}
Then the combination $\left[l/Z_l-(l-1)/Z_{l-1}\right]$ simplifies
to $ 2k/\!(\pi v_0b)$. Lastly, we notice that $\delta^0_l$ and
$\delta_{l-1}^ 0$ are related as \vspace{-0.3cm}
\begin{eqnarray}
\label{helpful} \delta^0_l-\delta^0_{l-1}=\pi/2.
\end{eqnarray} This follows from
Eq.~(\ref{asymtan}). Using the latter relation, it is easy to see
that all combinations containing phases in front of the square
brackets in the sum Eq.~(\ref{final}) are equal to $1$. Thus, the
sum over $l$ in Eq.~(\ref{final}) is simply equal to $2k^2/(\pi
v_0)$, and the result for $I^a(E)$ reads
\begin{eqnarray}\label{intermediate}
I^{a}(E)\Big |_{1\ll kb\ll \sqrt{v_0}b}=\frac{8\lambda } {\pi
b}kb.
\end{eqnarray}
We see that the two asymptotes Eq.~(\ref{smallkb}) and Eq.~(\ref{intermediate})
match at $kb\sim 1$. In both limits $I^{a}(E)$ increases essentially as $\propto E^{1/2}$
with $E$. Curiously, the magnitude
of potential, $v_0$, {\em drops out} from the
asymmetric part of the scattering rate in both limits.

The growth of $I^{a}(E)$ with energy is terminated
at $E\sim V_0$, i.e., at $kb\, \sim \, \nu b$, which also belongs to domain (ii). This
can be seen from Eqs. (\ref{0phase1}), (\ref{bigzl})
in the following way. For $l\sim \, kb \, \sim \, \nu b \gg 1$
we can use the asymptotic expression $\bigl[I_l^{\prime}(z)/I_l(z)\bigr]\approx
\left(1+l^2/z^2\right)^{1/2}$ for the ratio $I_l^{\prime}(\nu b)/I_l(\nu b)$.
Then Eq.~(\ref{bigzl}) yields
\begin{eqnarray}\label{Asymz1}
Z_l=\frac{\pi kb}{2}\left[1+\left(\frac l{kb}-\frac{\nu}{k}\sqrt{1+\frac{l^2}{\nu^2b^2}}\,\,
\right)^2\right].
\end{eqnarray}
Using the same asymptotic form in the expression
Eq.~(\ref{0phase1}) for phases, $\delta_l^0$ amounts to
replacement of $\nu/k$ by
$\left(\sqrt{\nu^2b^2+l^2}-l\right)/(kb)$ in the argument of the
arctangent in Eq.~(\ref{asymtan}). Still, the difference
$(\delta_{l}^0-\delta_{l-1}^0)$ remains $\pi/2$ with accuracy of a
small parameter $1/(kb)$. This is sufficient to replace by $1$ all
combinations containing phases in front of the square brackets in
the sum Eq.~(\ref{final}). Upon substituting Eq.~(\ref{Asymz1})
into Eq.~(\ref{final}), the summation over $l$ can be easily
performed, and we obtain
\begin{eqnarray}\label{satur}
&&I^{a}(E)=\frac{4\lambda v_0}k\sum_{l=1}^{kb}\bigl[l/Z_l-(l-1)/Z_{l-1}\bigr]\\
&&=\frac{4\lambda v_0}k\frac{kb}{Z_{kb}}=\frac{8\lambda
\sqrt{v_0}}\pi\frac{\sqrt{v_0}/k}
{1+\left(1-\sqrt{v_0}/{k}\right)^2}.\nonumber
\end{eqnarray}
Now we see that $I^a(E)$ reaches a maximum at $k =\left(v_0/2\right)^{1/2}$,
which corresponds to $E=V_0/2$. The maximal value is equal to
\begin{eqnarray}\label{maximal}
I^a(V_0/2)=\frac{4(\sqrt{2}+1)}{\pi}\lambda\sqrt{v_0}.
\end{eqnarray}
For smaller energies, Eq.~(\ref{satur}) reproduces
the result Eq.~(\ref{intermediate}).
Although Eq.~(\ref{satur}) was derived for
$E<V_0$, it remains applicable also for above-barrier scattering, $E>V_0$.
It is seen that, after passing the maximum, $I^a(E)$ falls off as $E^{-1/2}$.
 The energy dependence
of $I^a(E)$ is illustrated in Fig.~2.
%%%%%%%%%%%%%%%%%%%
\begin{figure}[t]
\centerline{\includegraphics[width=70mm,angle=0,clip]{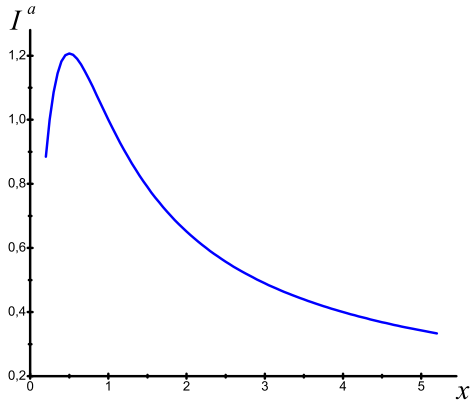}}
\caption{ (Color online) Intermediate energy domain:
$\hbar^2/(mb^2)\ll E\sim V_0 \ll  mV_0^2b^2/\hbar^2$. The
skew-scattering contribution, $I^a(E)$, to the scattering rate, in
the units $8\lambda\sqrt{v_0}/\pi$, is  plotted from
Eq.~(\ref{satur}) versus dimensionless energy $x=E/V_0$.}
\end{figure}
%%%%%%%%%%%%%%%%%%

We note that the position of maximum in $I^a(E)$
is model-dependent, in the sense,
that for a general potential with radius, $\sim b$, and magnitude,
$\sim V_0$, the position of maximum can differ from $V_0/2$ by a
numerical factor $\sim 1$. However, the existence of maximum in $I^a(E)$,
followed by $\propto E^{-1/2}$ decrease,  is model-independent.
Within a scaling factor, Eq.~(\ref{satur}) applies
up to the boundary,  $E\sim mV_0^2b^2/\hbar^2$,  of the domain (iii),
where the Born
approximation applies. From $E\sim V_0$ to this boundary $I^a(E)$
drops by a large factor $\sim \sqrt{v_0b^2}$.
In the high-energy tail (iii),  the behavior of $I^a(E)$
depends strongly on the shape of the
potential.

\section{$I^a(E)$ in the presence of quasilocal states}

The simplest model in which the quasilocal states emerge, is the
repulsive scattering potential with attractive core, as shown in
Fig. 1. Quasilocal state does not affect the scattering process
when the deviation of energy of the incident electron from the
resonance exceeds the width, $\Gamma$, of the quasilocal state.
For the model potential Fig. 1 the calculation of the width (with
prefactor) is presented in the Appendix. Below we consider  skew
scattering near the resonance for two particular cases:

\noindent 1. Resonance for zero angular momentum, $l=0$. In this
case, the phase $\delta_0(E)$ changes by $\pi$ as $E$ is swept
across the resonance. As we have seen above, in the low-energy
limit, $kb\ll1$, the phases $\delta_l^0$ fall off rapidly with
$l$. In calculating $I^a$, the phases, $\delta_{\pm 1}(E)$, should
be retained, since there is no skew scattering without them. All
phases $\delta_{l}(E)$ with $|l|\geq 2$ can be neglected.

\noindent 2. Resonance for angular momenta, $l=\pm 1$.
Now the phases $\delta_{\pm 1}(E)$ exhibit resonant behavior.
One has to retain the phase $\delta_0(E)$; all phases $\delta_{l}(E)$ with
$|l|\geq 2$ can be neglected.

Cases 1 and 2 are dramatically different because ${\hat H}_{so}$
{\em does not} split the level $l=0$, but {\em does} split levels
$l=\pm 1$. For this reason, in the case 1, $\delta^{1}_{\pm 1}(E)$
can be calculated perturbatively from Eq. (\ref{perturbative}), so
that for $\delta_0$ we have
\begin{eqnarray}\label{nearresonance0}
\delta_{0}=- \arctan\left[\frac{\Gamma_0}{2(E-E_0)}\right].
\end{eqnarray}
In the case 2,  $\delta_0$ is non-resonant and is still not affected by
${\hat H}_{so}$, while  {\em both} scattering
phases $\delta_1$ and $\delta_{-1}$ exhibit a resonant behavior
\begin{eqnarray}\label{nearresonance}
\delta_{1,\sigma}=\delta_{-1,-\sigma}=-
\arctan\left[\frac{\Gamma_1}{2(E-E_1-\delta E_{1,\sigma})}\right].
\end{eqnarray}
Now we have to express the splitting $\delta E_{1,\pm\sigma}$
through $\hat H_{so}$.
\begin{eqnarray}\label{splitting}
\delta E_{1,\sigma}\equiv \sigma\delta E_1=\lambda\sigma l
\frac{\int_0^{\infty}d\rho \frac{dV(\rho)}{d\rho}
\bigl\{R^0_1(\rho,E_1)\bigr\}^2}
{\int_0^{\infty}d\rho \rho \bigl\{R^0_1(\rho,E_1)\bigr\}^2},
\end{eqnarray}
where $R^0_1(\rho,E)$ is the radial wave function of the
{\em discrete} state, i.e., the coupling to continuum
is neglected. Important is that the {\em width}, $\Gamma_1$,
of the state $E_1$ is unaffected by the spin-orbit interaction.

Now we evaluate Eq.~(\ref{intrate}) for the cases 1 and 2. Upon
neglecting $\delta_l$ with $|l|\geq2$, we obtain from Eq.(\ref{SR})
a simplified form of $I^a(E)$
\begin{eqnarray}\label{inter}
%&&\hspace{-1cm}
I^a(E)\!\!&=&\!\!\frac\sigma{\pi
k}\bigl[ \sin(2\delta_{1,\sigma})+\sin2(\delta_0-\delta_{1,\sigma})\\
&&+\sin2(\delta_{-1,\sigma}-\delta_0)-\sin(2\delta_{-1,\sigma})\bigr]\nonumber\\
&=&\!\!\frac4{\pi
k}\sin(\delta_1-\delta_{-1})\sin(\delta_0-\delta_1-\delta_{-1})
\sin\delta_0.\nonumber
\end{eqnarray}
Since $I^a$ is independent of $\sigma$, in the last identity we had
dropped the subindex, $\sigma$, in the scattering phases.
%%%%%%%%%%%%%%%%%%
\begin{figure}[t]
\centerline{\includegraphics[width=80mm,angle=0,clip]{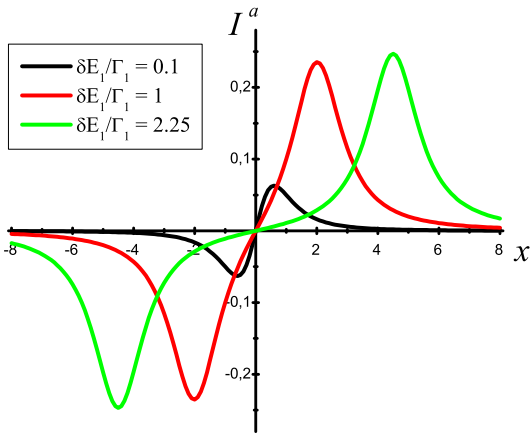}}
\caption{ (Color online) The shape of $I^a$ for small $E, E_1\ll
\hbar^2/(mb^2)$, is plotted from Eq.~(\ref{second}) versus
dimensionless deviation $x=2(E-E_1)/\Gamma_1$ from the resonance,
for three different values of dimensionless spin-orbit splitting,
$\delta E_1/\Gamma_1$. }
\end{figure}
%%%%%%%%%%%%%%%%%%
In the case 1, the phases $\delta_1$ and  $\delta_{-1}$ are small,
and Eq.~(\ref{inter}) assumes the form
\begin{eqnarray}\label{first}
I^a(E)\!\!&=&\!\!\frac4{\pi k}\left(\delta_1-\delta_{-1}\right)\sin^2\delta_0\\
&\approx&\!\! \frac1{\pi
k}(\delta_1-\delta_{-1})\frac{\Gamma^2_0}{(E-E_0)^2+\Gamma^2_0/4}.\nonumber
\end{eqnarray}
The difference $\left(\delta_1-\delta_{-1}\right)$ is proportional
to spin-orbit constant, $\lambda$, and can be evaluated from Eqs.
(\ref{viaz}) and (\ref{ZL}), namely,
\begin{eqnarray}
\delta_1-\delta_{-1}=
\frac{\pi^3\lambda}{4b^2}(kb)^2.
\end{eqnarray}
Eq.~(\ref{first}) indicates
that in the case 1 the energy dependence of the skew scattering is
the same as the energy dependence of $\sigma^s(E)$, i.e., it
exhibits a resonant enhancement near $E=E_0$.

In the case 2, $\delta_0$ is small and nonresonant, while
$\delta_1$ and $\delta_{-1}$ have resonances at $E=E_1\pm \delta
E_1$. Upon using Eq.~(\ref{nearresonance}), we find
%\begin{widetext}
\begin{eqnarray}\label{second}
%&&\hspace{-1cm}
&&\hspace{-0.3cm}I^a(E)\approx-\frac{4\delta_0}{\pi k}\\
&&\hspace{-0.3cm}\times\frac{\Gamma_1^2(E-E_1)\delta E_1}{\left[(E-E_{1}-\delta
E_1)^2+\Gamma^2_1/4\right] \left[(E-E_{1}+\delta
E_1)^2+\Gamma^2_1/4\right]}.\nonumber
\end{eqnarray}
%\end{widetext}
Our prime observation is that, in the case 2, the behaviors of
$\sigma^s(E)$ and $I^a(E)$ are vastly different. While
$\sigma^s(E)$ is $\propto
\Gamma_1^2/\left[(E-E_1)^2+\Gamma_1^2/4\right]$, the skew
scattering is an {\em odd} function of $(E-E_1)$. Actual shape of
$I^a(E)$ is determined by the relation of two small energies
$\delta E_1$ and $\Gamma_1$. For  $\delta E_1\ll \Gamma_1$,
position of a maximum (and minimum) and the width of $I^a(E)$ are
all $\sim \Gamma_1$. In the opposite limit,  $\delta E_1\gg
\Gamma_1$, the extrema are located at $(E-E_1)=\pm \delta E_1$,
while the width $\sim \Gamma_1$ is much smaller. Behavior of the
skew scattering near the resonance is illustrated in Fig.~3 for
both limits.

At this point we emphasize a significant difference between the
two cases.  In the  case 1 we have $I^a\propto dV/d\rho$ at
$\rho=b$ while in the case 2, the characteristic scale $\delta
E_1$ is $\propto dV/d\rho$ at $\rho=a$.

In the end of this Section we consider the case when the resonance
$E=E_1$ falls into the domain of ``semiclassical'' scattering,
$1\ll kb \lesssim v_0^{1/2}b$. Nonresonant skew scattering is
described by Eq.~(\ref{intermediate}) in this domain and requires
taking into account all the angular momenta up to $l=kb$. For the
same reason, in calculating resonant skew scattering, one has to
keep all four terms in the sum Eq.~(\ref{SR}) which contain either
$\delta_1$ or $\delta_{-1}$. We have
\begin{widetext}
\begin{eqnarray}\label{interhigh}
&&\hspace{-0.3cm}
\delta I^a_{\mbox{r}}(E)\!=\!\frac1{\pi k}
\bigl[\sin2(\delta_{1}-\delta_{2})+\sin2(\delta_0-\delta_{1})
+\sin2(\delta_{-1}-\delta_0)+\sin2(\delta_{-2}-\delta_{-1})\bigr]
\\
&&=\!\frac2{\pi k}\bigl[\sin(\delta_1-\delta_{-1}-\delta_2+\delta_{-2})
-\sin(\delta_1-\delta_{-1})\bigr] \cos(\delta_1+\delta_{-1}-2\delta_0)
=(\delta^1_{-2}-\delta^1_{2})\bigl[\cos2(\delta_1-\delta_{0})+
\cos2(\delta_{-1}-\delta_{0})\bigr].\nonumber
\end{eqnarray}
\end{widetext}
In the second identity we used the fact that
$\delta_2+\delta_{-2}\approx 2\delta^0_0$, which follows from
Eq.~(\ref{asymtan}); in the third identity we replaced
$\sin(\delta_2-\delta_{-2})$, which is nonzero only due to
spin-orbit-induced corrections, by
$\left(\delta_2^{1}-\delta_{-2}^1\right)\approx 4\lambda k/b$. We
note now, that the nonresonant parts of differences
$(\delta_1-\delta_0)$ and $(\delta_{-1}-\delta_0)$ are $\pi/2$ and
$-\pi/2$, respectively [see Eq.~(\ref{helpful})], while the
resonant parts of $\delta_1$ and $\delta_{-1}$ are given by
Eq.~(\ref{nearresonance}). This leads us to the final result
\begin{eqnarray}\label{kak2kak} \delta
I^a_{\mbox{r}}(E)\!\!&\approx&\!\! \frac{4\lambda}{\pi b}
\left\{ \frac{(E-E_{1}-\delta E_{1})^2
-\Gamma_1^2/4}{(E-E_{1}-\delta E_{1})^2+\Gamma_1^2/4}\right.\\
&&\hspace{1.5cm}\left.+ \frac{(E-E_{1}+\delta E_{1})^2
-\Gamma_1^2/4} {(E-E_{1}+\delta E_{1})^2+ \Gamma_1^2/4}
\right\}.\nonumber
\end{eqnarray}
It is seen from Eq. (\ref{kak2kak}) that the characteristic
 magnitude of the resonant skew scattering contribution
is $\delta I^a_{\mbox{r}}(E)\sim \lambda/b$. On the other hand,
the background value,  $I^a(E)\sim \lambda k$
[see Eq.~(\ref{intermediate})], is much larger.
This is because the background value is the sum
of $kb \gg 1$ contributions.
Although $\delta I^a_{\mbox{r}}(E)$ constitutes a small
correction, it has a lively energy dependence.
This dependence is illustrated in Fig.~4
for different values of dimensionless ratio
$\delta E_1/\Gamma_1$. We see that, as $\delta E_1/\Gamma_1$
increases, the structure in $\delta I^a_{\mbox{r}}(E)$ crosses
over from one minimum to two minima.
%%%%%%%%%%%%%%%%%%
\begin{figure}[t]
\centerline{\includegraphics[width=80mm,angle=0,clip]{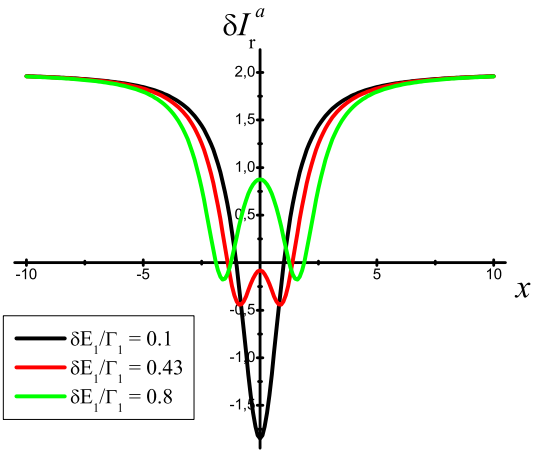}}
\caption{ (Color online) The shape of the resonant contribution to
the skew-scattering part of the scattering rate, $\delta
I^a_{\mbox{r}}$, in the domain, $\hbar^2/(mb^2)\ll E \lesssim
V_0$, is plotted from Eq.~(\ref{kak2kak}) versus dimensionless
deviation, $x=2(E-E_1)/\Gamma_1$, for three different values of
dimensionless ratio $\delta E_1/\Gamma_1$. }
\end{figure}
%%%%%%%%%%%%%%%%%%

\section{Many-body effects in the resonant skew scattering}

As it was mentioned in the Introduction, the quantity relevant for
transport is $I^a(E)$ at $E=E_F$. Then the resonance condition,
$E_F=E_0$, can be satisfied either if  electron concentration or
the potential of the attractive core can be controlled. On the
other hand, even away from resonance, at $E_F>E_0$, the symmetric
part of scattering cross section, $\sigma^s(E_F)$, experiences a
strong enhancement when the temperature, $T$, is lower than the
Kondo temperature, $T_K$. This prominent many-body effect stems
from the Hubbard repulsion of two electrons in the quasilocal
level.

Spin-orbit interaction does not lift the degeneracy
of the ground state, and thus, does  not affect $\sigma^s(E_F)$.
On the other hand, since the skew scattering is essentially due
to interference of $l=0$ and $l=\pm 1$ scattering channels,
one should expect the Kondo
enhancement of $\sigma^a$ alongside with enhancement of
$\sigma^s$.

Quantitatively, this enhancement can be found
in the following way.
Kondo effect modifies Eq.~(\ref{first}). Namely,
the scattering phase, $\delta_0$, should be
replaced by the temperature-dependent Kondo
phase, $\delta_0(T)$. The low-$T$ and high-$T$
asymptotes of $\delta_0(T)$ are the following
\begin{eqnarray} \label{Kr}&&\hspace{-.6cm}
\sin^2\delta_0= \left\{
\begin{array}{l}
1-\pi^2\left(\frac T{T_K}\right)^2,\quad T\ll T_K,\\
\,\\
\frac {3\pi^2} {16}\frac1{\ln^2\left(T/T_K\right)},\quad T\gg T_K,
\end{array}
\right.
\end{eqnarray}

Eq.~(\ref{Kr}) describes how $I^a$ drops with increasing
temperature from its ``unitary'' value $I^a= \pi^2\lambda k_F$,
where $k_F$ is the Fermi momentum, to the background value, given
by Eq.~(\ref{smallkb}) with $k\rightarrow k_F$.

A very lively behavior of the skew scattering emerges when the
quasilocal level, $E_1$, is close to the Fermi level, while $E_0$
is well below the Fermi level. Then the Kondo resonance develops
in $l=0$ channel at $T< T_K$, while the phases $\delta_{\pm
1}(E_F)$ are strongly sensitive to the deviation $(E_1-E_F)$.

A particularly interesting issue is what happens to the sign
reversal of $I^a$, found in the previous Section. In this case,
both $\delta_0$ and $\delta_{\pm1}$ are resonant; $\sin\delta_0$
is defined by Kondo resonance Eq.~(\ref{Kr}), whereas
$\delta_{\pm1}$ are given by Eq.~(\ref{nearresonance}). Upon
setting $E=E_F$ in Eq.~(\ref{nearresonance}) and substituting it
into Eq. (\ref{inter}), we obtain
\begin{widetext}
\begin{eqnarray}\label{kondo+}
I^a(E_1, E_F,T)
\approx\frac{4\delta E_1\Gamma_1}{\pi k}\sin^2\delta_0(T)
\frac{\Gamma_1(E_F-E_1)\cot\delta_0(T)+\left[ (E_F-E_{1}-\delta
E_1)(E_F-E_{1}+\delta E_1)-\Gamma_1^2/4\right]}
{\bigl[(E_F-E_{1}-\delta E_1)^2+\Gamma_1^2/4\bigr]
\bigl[(E_F-E_{1}+\delta E_1)^2+\Gamma_1^2/4\bigr]}.
\end{eqnarray}
\end{widetext}
At ``high'' temperature, $\delta_0$ is small and we return to
Eq.~(\ref{second}). However, deep in the Kondo regime, the
behavior of $I^a(E_F)$ changes drastically. As it is seen from
Eq.~(\ref{kondo+}),  the dominant contribution to $I^a$ comes from
the second term in the numerator, which is {\em even} function of
$(E_F-E_1)$, unlike the first term, which is odd. Therefore, upon
decreasing temperature, the curve, $I^a(E_F)$, evolves from
asymmetric to symmetric.

For $\delta_0=\pi/2$, the shape of $I^a$ versus $E_F$, which is
proportional to the electron density, is shown in Fig.~5. We see
that symmetric shape undergoes a strong transformation as the
splitting $\delta E_1/\Gamma_1$ increases. For each value of
splitting there are {\em two} points of the sign change positioned
symmetrically with respect to the point $E_F=E_1$.

\section{Concluding remarks}
The Lorentzian shape of $I^a(E)$, described by Eq. (\ref{first}),
suggests that it should be accompanied by the Fano feature
\cite{Fano}, since $I^a(E)$ must assume its nonresonant value away
from $E=E_0$. The issue of Fano resonance near $E=E_1$ is more
delicate. As can be seen from Eq. (\ref{second}) at the
Fano-resonance condition $\delta_1\approx \delta_{-1}\approx
\delta_0/2$ the factor $\sin(\delta_1-\delta_{-1})$ turns to zero.
This observation can be interpreted as ``cancellation'' of Fano
resonances due to interference in $l=1$ and $l=-1$ channels.
Similar ``cancellation'' was pointed out in
Refs.~\onlinecite{Dm1,Dm2}. In these papers the photo-current,
caused by infrared excitation of electron either from impurity
into the conduction band \cite{Dm1} or between two Zeeman
subbands\cite{Dm2}  of $InSb$ in a strong magnetic field, was
studied  experimentally. In fact, there is a general similarity
between the sine reversal of the skew scattering with energy and
the sign reversal of photocurrent as a function of magnetic field,
observed in Refs.~\onlinecite{Dm1,Dm2}, the underlying reason
being the interference of a non-resonant and {\em two} split
resonant channels.

%%%%%%%%%%%%%%%%%%
\begin{figure}[t]
\centerline{\includegraphics[width=80mm,angle=0,clip]{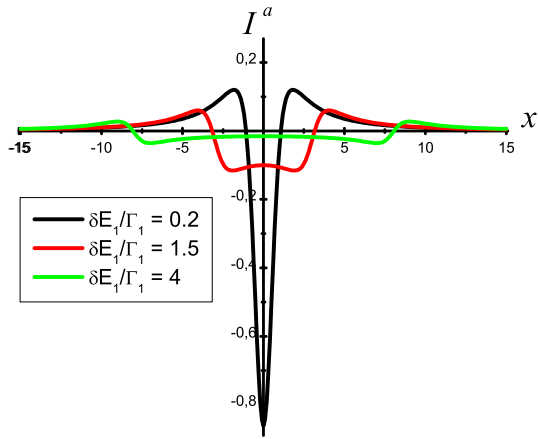}}
\caption{ (Color online) The shape of $I^a$ in the Kondo regime,
$T\ll T_K$, and resonance with quasilocal level, $E=E_1$, is
plotted from Eq.~(\ref{kondo+}) versus dimensionless deviation
$x=2(E_1-E_F)/\Gamma_1$ of $E_1$ from the Fermi level for three
values of the dimensionless spin-orbit splitting, $\delta
E_1/\Gamma_1$, of the level $E=E_1$.}
\end{figure}
%%%%%%%%%%%%%%%%%%

\section{ACKNOWLEDGEMENT}
This work was supported by NSF under Grant No. DMR-0503172.
\appendix

\section{}

In this Appendix we derive an analytical expression for the width
of a 2D quasilocal state. The Schr\"{o}dinger equation for the
radial part of the wave function, $R_{l}(\rho, E)$, in a
azimuthally-symmetric potential, $V(\rho)$, reads
\begin{eqnarray}\label{Sche}
R_{l}^{\prime\prime} + \frac1\rho R_{l}^{\prime}
+\left(\frac{2m}{\hbar^2}\Bigl[E_l-V(\rho)\Bigr]-\frac{l^2}{\rho^2}\right)R_{l}=0.
\end{eqnarray}
Upon substitution
\begin{eqnarray}\label{subst}
R_{l}(\rho, E)=\frac{\chi_{l}(\rho, E)}{\sqrt{\rho}},
\end{eqnarray}
Eq.~(\ref{Sche}) acquires a Hermitian form
\begin{eqnarray}\label{uSche}
\chi_{l}^{\prime\prime} + \left(\frac{2m}{\hbar^2}
\Bigl[E_l-V(\rho)\Bigr]-\frac{l^2-1/4}{\rho^2}\right)\chi_{l}=0.
\end{eqnarray}
Consider now the potential, depicted in Fig.~1 with $V(\rho)=0$
for $\rho>b$. To calculate the width, $\Gamma$, we consider an
auxiliary potential, $\tilde{V}(\rho)$, which coincides with
$V(\rho)$ for $\rho<b$ and is a constant $\tilde{V}(\rho)=V_0$ for
$\rho>b$. Localized state, $E=\tilde{E}_l$ in this potential is
stationary. Denote with $\tilde{\chi}_l(\rho)$ the corresponding
radial wave function, so that
\begin{eqnarray}\label{uSche1}
\tilde{\chi}_{l}^{\prime\prime} + \left(\frac{2m}{\hbar^2}
\Bigl[\tilde{E}_l-\tilde{V}(\rho)\Bigr]-\frac{l^2-1/4}{\rho^2}\right)
\tilde{\chi}_{l}=0.
\end{eqnarray}
Upon multiplying Eq.~(\ref{uSche}) by $\tilde{\chi}_l$ and Eq.~(\ref{uSche1}) by $\chi_l$, subtracting
and integrating from $a$ to $b$ we obtain the relation
\begin{eqnarray}\label{mixed}
\bigl[\tilde{\chi}_l\chi_l^{\prime} -\chi_l
\tilde{\chi}_l^{\prime}\bigr]\Big|_a^b= \frac{2m}{\hbar^2}
(\tilde{E}_l-E_l)\!\int\limits_a^b\!\!d\rho\,\chi_l\tilde{\chi}_l.
\end{eqnarray}
The fact that the difference $(\tilde{E}_l-E_l)$ is much smaller
than $E_l$ allows us to keep in the left-hand side only the
contribution from $\rho=b$. Then it is convenient to rewrite Eq.
(\ref{mixed}) as
\begin{eqnarray}\label{mixed1}
\left[\frac{\chi_l^{\prime}}{\chi_l}(b) -\frac{
\tilde{\chi}_l^{\prime}}{\tilde{\chi}_l}(b)\right]=
\frac{2m(\tilde{E}_l-E_l)} {\hbar^2\chi_l(b)\tilde{\chi}_l(b)}
\int\limits_a^b\!\!d\rho\,\chi_l(\rho)\tilde{\chi}_l(\rho).
\end{eqnarray}
To find the imaginary part of $E_l$ we use the continuity of the
logarithmic derivatives of $\chi_l$  at $\rho=b$
\begin{eqnarray}\label{QC1}
%&&\frac{\tilde{\chi}^{\prime}_l}{\tilde{\chi}_l}(b)=\frac1{2b}+
%\nu\frac{K^\prime_l(\nu b)}{K_l(\nu b)},\\
&&\frac{\chi^\prime_l}{\chi_l}(b)=\frac1{2b}+
k\frac{H^{+\prime}_l(kb)}{H^+_l(kb)},
\end{eqnarray}
where $H^+_l(k\rho)$ is the Hankel function corresponding to
outgoing wave at $\rho \rightarrow \infty$. Since
$\tilde{\chi}_l(\rho)$ is real, the second term in the left-hand
side of Eq.~(\ref{mixed1}) does not contribute to $\Im m E_l$. The
imaginary part of $E_l$ originates from logarithmic derivative of
the Hankel function in Eq.~(\ref{QC1}). Final expression for
$\Gamma_l=\Im m E_l$ emerges upon setting
$\chi_l(\rho)=\tilde{\chi}_l(\rho)=\sqrt{\rho}K_l(\nu\rho)$, where
$K_l$ is the Macdonald function, which is the solution of
Eq.~(\ref{uSche1}) in the barrier region, and extending the upper
limit of integration in Eq.~(\ref{mixed1}) to infinity. Both steps
are justified if $\Im m E_l \ll E_l$. Then we obtain
\begin{eqnarray}\label{WC}
&&\hspace{-1.1cm}\Gamma_l(E)=\frac{\hbar^2}{\pi
m}\cdot\frac1{J_l^2(kb) +N_l^2(kb)}\cdot \frac{K_l^2(\nu
b)}{\int_a^\infty d\rho\rho K_l^2(\nu \rho)}.
\end{eqnarray}
Further simplification is achieved when the core is wide enough,
so that $\nu a =\left[2m(V_0-E_l)a^2\right]^{1/2}/\hbar$ is large.
Then we can use the large-$\rho$ asymptote of the Macdonald
function in the integrand in Eq.~(\ref{WC}), which yields

\begin{eqnarray}\label{WC1}
&&\hspace{-1cm}\Gamma_l(E)=\frac{2\hbar^2}{\pi m b}\cdot
\frac{\nu\, e^{-2\nu(b-a)}}{J_l^2(kb) +N_l^2(kb)}.
\end{eqnarray}
In the opposite limit, $\nu a \ll 1$, the exponent in the width, $\Gamma_l$, is
$\exp(-2\nu b)$, while the dependence on $a$ enters into the prefactor.

\end{document}